\documentclass[conference,10pt,letterpaper,twocolumn]{IEEEtran}

\usepackage[utf8]{inputenc}
\usepackage[T1]{fontenc}
\usepackage[htt]{hyphenat}  
\usepackage[pdftex]{graphicx}
\usepackage{amsmath}
\usepackage{url}
\usepackage{orcidlink}
\usepackage[inline]{enumitem}
\setlist[description]{font=\itshape\underline}
\usepackage{todonotes}
\usepackage{caption}
\usepackage{subcaption}
\usepackage{booktabs}
\usepackage{xcolor}
\usepackage[detect-weight=true,detect-family=true,binary-units=true,list-units=single,range-units=single]{siunitx}
\usepackage[nolist]{acronym}
\begin{acronym}
    \acro{CPS}{Cyber-Physical System}
    \acro{NCS}{Networked Control System}
    \acroindefinite{NCS}{an}{a}
    \acro{LAN}{Local-Area Network}
    \acro{WLAN}{Wireless Local-Area Network}
    \acro{KPI}{Key Performance Indicator}
    \acro{WPAN}{Wireless Personal Area Network}
    \acro{CSMA/CD}{Carrier-Sense Multiple Access with Collision Detection}
    \acro{CLEAVE}{ControL bEnchmArking serVice on the Edge}
    \acro{OS}{Operating System}
    \acro{UDP}{User Datagram Protocol}
    \acro{TCP}{Transmission Control Protocol}
    \acro{RMS}{Root Mean Square}
    \acro{RTT}{Round-Trip Time}
    \acro{CI}{Confidence Interval}
    \acro{AP}{Access Point}
    \acro{API}{Application Programming Interface}
    \acro{SSF}{Swedish Foundation for Strategic Research}
    \acro{TECoSA}{Trustworthy Edge Computing Systems and Applications}
    \acro{ABC}{Abstract Base Class}
    \acroplural{ABC}[ABCs]{Abstract Base Classes}
    \acro{URL}{Uniform Record Locator}
    \acro{AWS}{Amazon Web Services}
    \acro{EC2}{Elastic Compute 2}
    \acro{AMI}{Amazon Machine Image}
    \acro{SSH}{Secure Shell}
    \acro{IP}{Internet Protocol}
    \acro{VPN}{Virtual Private Network}
    \acro{NAT}{Network Address Translation}
    \acro{NTP}{Network Time Protocol}
    \acroindefinite{NTP}{an}{a}
    \acro{SDR}{Software-Defined Radio}
    \acroindefinite{SDR}{an}{a}
    \acro{VLAN}{Virtual Local-Area Network}
    \acro{APT}{Advaced Packaging Tool}
    \acro{LTE}{Long-Term Evolution}
    \acro{SSID}{Service Set Identifier}
    \acro{EPC}{Evolved Packet Core}
    \acro{UE}{User Equipment}
    \acro{eNodeB}{Evolved Node B}
    \acro{DNS}{Domain Name System}
    \acro{MNIST}{Modified National Institute of Standards and Technology}
    \acro{HTTP}{Hyper-Text Transfer Protocol}
    \acroindefinite{HTTP}{an}{a}
    \acro{YAML}{YAML Ain't Markup Language}
    \acro{O-RAN}{Open Radio Access Network}
    \acro{FOSS}{Free and Open Source Software}
    \acro{COSMOS}{Cloud enhanced Open Software defined MObile wireless testbed for city-Scale deployment}
    \acro{OMF}{ORBIT Management Framework}
    \acro{POWDER}{Platform for Open Wireless Data-driven Experimental Research}
    \acro{COTS}{Commercial Off-The-Shelves}
    \acro{RF}{Radio-Frequency}
    \acro{VM}{Virtual Machine}
    \acro{REST}{REpresentational Sate Transfer}
    \acro{OS}{Operating System}
    \acro{MaaS}{Metal-as-a-Service}
    \acro{OEDL}{\ac{OMF} Experiment Description Language}
    \acro{RSpec}{Resource Specification}
    \acro{LXC}{LinuX Containers}
    \acro{MEC}{Mobile Edge Computing}
    \acro{USRP}{Universal Software Radio Peripheral}
    \acro{OAI}{OpenAirInterface}
    \acro{NR}{New Radio}
    \acro{FPGA}{Field-Programmable Gate Array}
\end{acronym}

\usepackage[
  style=numeric-comp,
  sorting=none,
  sortcites,
  hyperref,
  mincitenames=1,
  maxcitenames=2,
  maxbibnames=2,
  minbibnames=1,
  citestyle=numeric-comp, 
  backend=bibtex
]{biblatex}
\bibliography{references.bib}
\AtBeginBibliography{\small}
\AtEveryBibitem{\clearfield{day}}
\AtEveryBibitem{\clearfield{isbn}}
\AtEveryBibitem{\clearfield{series}}
\AtEveryBibitem{\clearlist{location}}
\AtEveryBibitem{\clearfield{doi}}

\usepackage{outline}
\usepackage{orcidlink}
\usepackage[all]{hypcap}
\usepackage{makecell}
\usepackage[capitalize,nameinlink,noabbrev]{cleveref}

\hypersetup{
  hidelinks,
  colorlinks=true,
  allcolors=black,
  pdfstartview=Fit,
  breaklinks=true
}

\newcommand{\kthemail}[1]{\href{mailto:#1@kth.se}{#1}}

\begin{document}
\title{Ainur:~A Framework for Repeatable End-to-End Wireless Edge Computing Testbed Research}

\author{%
  \IEEEauthorblockN{%
  Manuel {Olguín Muñoz}\IEEEauthorrefmark{1}~\orcidlink{0000-0002-3383-2335}, Seyed Samie Mostafavi\IEEEauthorrefmark{2}~\orcidlink{0000-0001-9316-0414}, Vishnu N. Moothedath\IEEEauthorrefmark{3}~\orcidlink{0000-0002-2739-5060}, James Gross\IEEEauthorrefmark{4}~\orcidlink{0000-0001-6682-6559}%
  }%
  \IEEEauthorblockA{%
    School of Electrical Engineering \& Computer Science\\%
    KTH Royal Institute of Technology, Sweden\\%
    Email:~\{\IEEEauthorrefmark{1}\kthemail{molguin}, \IEEEauthorrefmark{2}\kthemail{ssmos}, \IEEEauthorrefmark{3}\kthemail{vnmo}, \IEEEauthorrefmark{4}\kthemail{jamesgr}\}\href{mailto:molguin@kth.se}{@kth.se}
  }
}
\maketitle

\begin{abstract}
  Experimental research on wireless networking in combination with edge and cloud computing has been the subject of explosive interest in the last decade.
  This development has been driven by the increasing complexity of modern wireless technologies and the  extensive softwarization of these through projects such as a \acf*{O-RAN}.
  In this context, a number of small- to mid-scale testbeds have emerged, employing a variety of technologies to target a wide array of use-cases and scenarios in the context of novel mobile communication technologies such as 5G and beyond-5G.
  Little work, however, has yet been devoted to developing a standard framework for wireless testbed automation which is hardware-agnostic and compatible with edge- and cloud-native technologies.
  Such a solution would simplify the development of new testbeds by completely or partially removing the requirement for custom management and orchestration software.

  In this paper, we present the first such mostly hardware-agnostic wireless testbed automation framework, \emph{Ainur}.
  It is designed to configure, manage, orchestrate, and deploy workloads from an end-to-end perspective.
  Ainur is built on top of cloud-native technologies such as Docker, and is provided as \acs*{FOSS} to the community through the {KTH-EXPECA/Ainur} repository on GitHub.
  We demonstrate the utility of the platform with a series of scenarios, showcasing in particular its flexibility with respect to physical link definition, computation placement, and automation of arbitrarily complex experimental scenarios.
\end{abstract}
\section{Introduction}
Experimental research in the area of wireless networking has received ever increasing attention over the last years, driven, on the one hand, by the complexity of modern networked systems and corresponding applications. 
On the other, networked systems are more and more based on software instead of dedicated hardware, which allows experimental testbeds to be rededicated simply through an update as system versions evolve --- in contrast to the redeployment of hardware necessitated \numrange[range-phrase={--}]{10}{15} years ago.
The complexity of these systems, as well as their softwarization are expected to continue growing, driving in turn an expanding interest in testbed-based experimental research in wireless systems.

Over the last years, several small- to mid-scale testbeds have emerged that leverage a large degree of freedom with respect to hardware and software, for example the 
\begin{enumerate*}[itemjoin={{, }}, itemjoin*={{, and }}]
    \item \acs*{COSMOS}
    \item \acs*{POWDER}
    \item Drexel Grid \ac{SDR}
\end{enumerate*} testbeds.
\acs{COSMOS}~(\emph{\acl{COSMOS}})\acused{COSMOS} is a testbed spanning an area of roughly \num{1} square mile (\SI{2.6}{\kilo\meter\squared}) featuring \acp{SDR}, \si{\milli\meter}-wave equipment, optical fibers, cloud integration, and compute for core network functionality and application data processing~\cite{Cosmos1,cosmos2}.
It contains over \num{200} rooftop, intermediate, and mobile nodes, and is controlled and managed by a central node.
\ac{COSMOS} relies on the \ac{OMF} (originally developed for ORBIT~\cite{orbit}), and employs the \ac{OEDL}, a domain-specific imperative language based on Ruby, for experiment development and definition.

\acs{POWDER}~(\emph{\acl{POWDER}})\acused{POWDER} promises research on wireless and mobile networks with a level of programmability down to the waveform~\cite{powder}.
The testbed spans a \SI{15}{\kilo\meter\squared} area and features about \num{15} fixed programmable radio nodes, based on off-the-shelves \acp{SDR} and featuring edge-like compute capabilities and integration with cloud resources, which interact with \num{50} mobile nodes. 
\ac{POWDER} experiments are defined and developed in \emph{profiles}, which correspond to \ac{VM} images containing the necessary software and configurations.
These profiles are defined through using \ac{RSpec}\footnote{\url{https://groups.geni.net/geni/wiki/GENIExperimenter/RSpecs}} documents.

Finally, the \emph{Drexel Grid \ac{SDR} Testbed} features \acp{SDR} that connect either over-the-air, through a channel emulator, or over a combination of the two, to facilitate realistic and reproducible experimentation~\cite{DrexelGrid}.
Primarily intended for \ac{SDR}-centric research, it does not integrate any core, cloud or edge components.
However, the testbed extensively employs the \ac{LXC} runtime for the deployment of both experimental code and \ac{SDR} software, which affords users great freedom when it comes to development of experiments.

Experimentation is key to to fully understanding the implications of next-generation wireless systems, cloud, and edge computing paradigms, and thus more of these testbeds are sure to emerge in the near future.
Yet, little work has so far been devoted to general-purpose, hardware-agnostic software frameworks for the management and automation of such systems.
Existing platforms implement their own, ad-hoc software solutions which are not compatible with other testbeds, and in many cases are not even compatible with reigning cloud-native standards.
This is, for instance, the case with \ac{COSMOS} and \ac{POWDER}; their reliance on domain-specific languages limits their integration with cloud-native solutions, which generally build upon general-purpose languages such as Python and Go.
These testbeds further leverage virtualization technology based on \acp{VM} instead of more lightweight and edge-compatible solutions such as containers.

To the best of our knowledge, CloudRAFT~\cite{cloudraft} is the only work to tackle (to a certain extent) this challenge.
CloudRAFT corresponds to a cloud-based framework for mobile network experimentation, with a focus on simplifying the management of testbed resources.
The goal of this project is to integrate, coordinate, share, and improve upon existing testbeds, and employs pre-built \acp{VM} containing the necessary software for experiments.
Although it provides some automation for testbed resource provisioning and experiment execution, its focus is largely rather on the sharing and partitioning of testbed systems.
Testbeds currently working with CloudRAFT include a variety of domain-specific setups, including \iac{SDR}-based testbed as a well as a ground vehicular robot for mobility-related experimentation.

In this work, we present our solution to the challenge of testbed automation: Ainur, a framework for wireless testbed automation with a specific focus on end-to-end experimental research in the context of edge-computing using cloud- and edge-native technologies.
Ainur is designed to deploy experimental runs from a workload perspective by configuring the physical testbed, initializing all involved software components, deploying and executing the experimental workload, collecting logs and data, and finally gracefully degrading the system.
The framework allows for dynamic, software-definition of physical and logical links, network topology, cloud and edge computing resources, as well as experimental workload deployment and orchestration.
It heavily leverages cloud-native technologies, such as Docker containers, in order to supports a wide variety of different testbed hardware setups and experimental configurations and workloads, as well as to be as easily extendable as possible.
Furthermore, we make Ainur available to the community as \ac{FOSS}.
It can be obtained from the {KTH-EXPECA/Ainur} repository on GitHub~\cite{ainur:github}, released under an Apache version \num{2.0} license.

The rest of this paper is structured as follows.
\cref{sec:ainur} presents the framework as well as the key concepts and technologies supporting its design and architecture.
This section also discusses briefly the assumptions made about the underlying hardware on which the framework is set to run, and we present an overview of our experimental testbed.
Next, \cref{sec:demo} describes two demonstration procedures through which we will showcase the flexibility and potential of this tool for the automatic, repeatable, end-to-end experimentation in the context of wireless testbeds.
Finally, in \cref{sec:conclusion} we summarize our contributions and conclude this paper.






\section{The Ainur Framework}\label{sec:ainur}

\begin{figure}[t]
    \centering
    \includegraphics[width=0.9\linewidth]{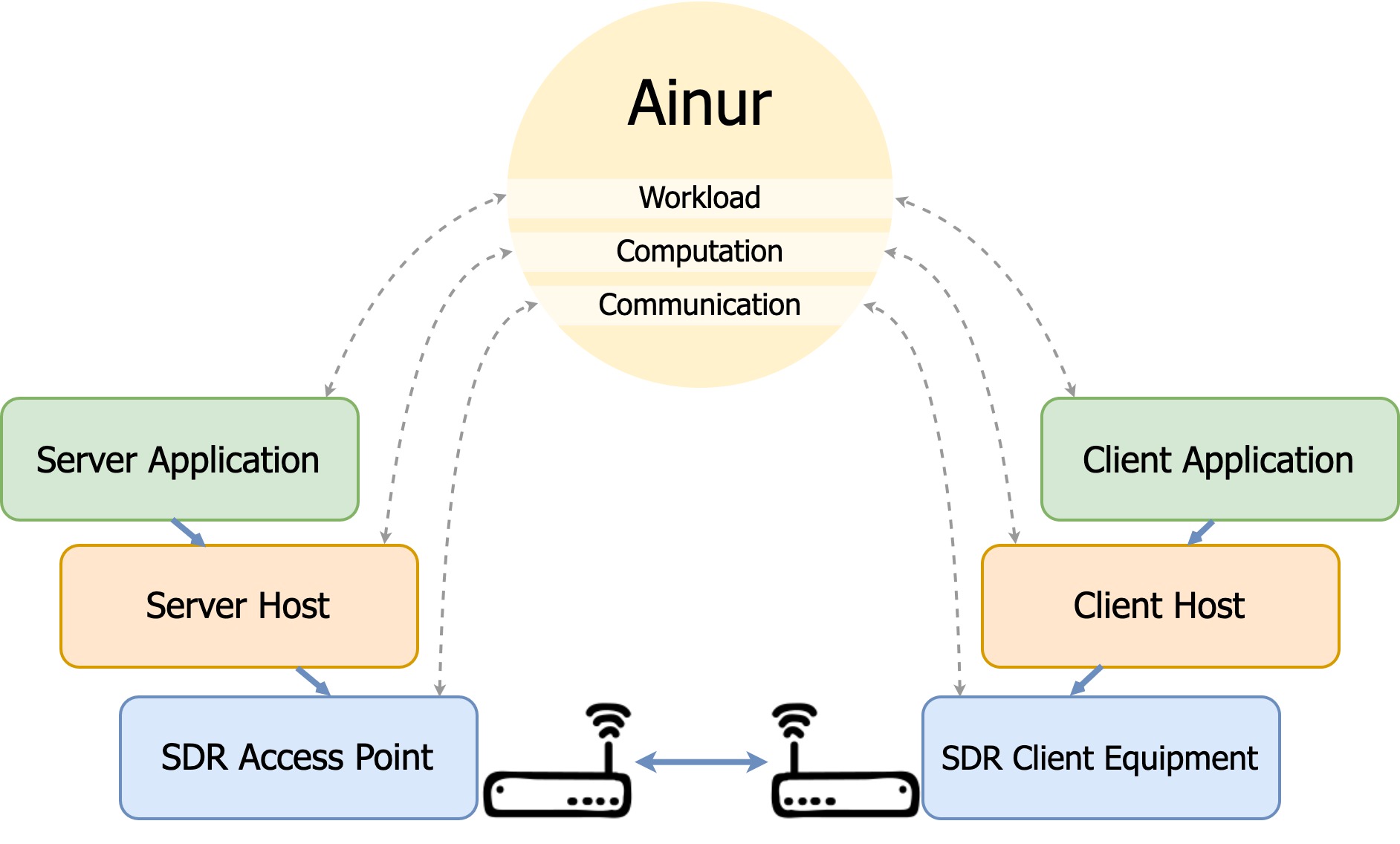}
    \caption{Layered structure of an Ainur experiment}\label{fig:overview}
\end{figure}

\begin{figure*}
    \centering
    \includegraphics[width=0.8\textwidth]{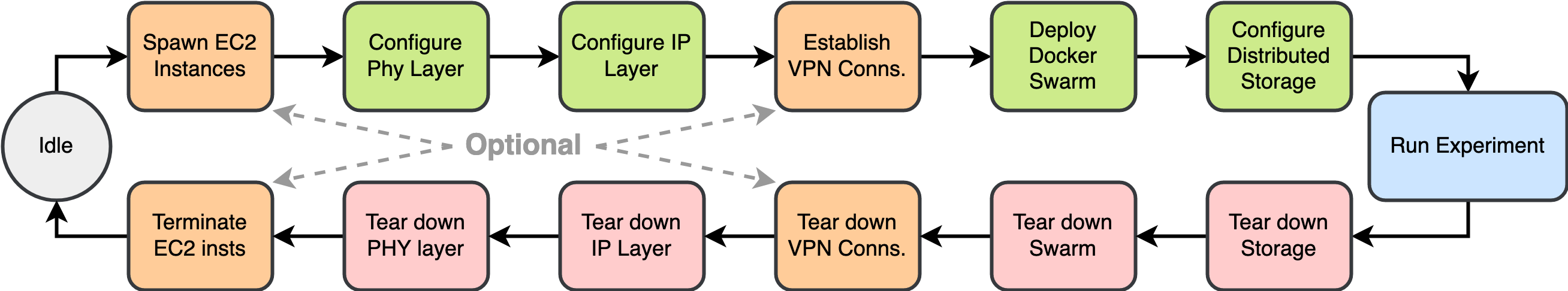}
    \caption{Lifecycle of an experimental run in Ainur. Blocks in orange are optional.}\label{fig:flow}
\end{figure*}

\begin{figure*}
    \centering
    \includegraphics[width=.8\textwidth]{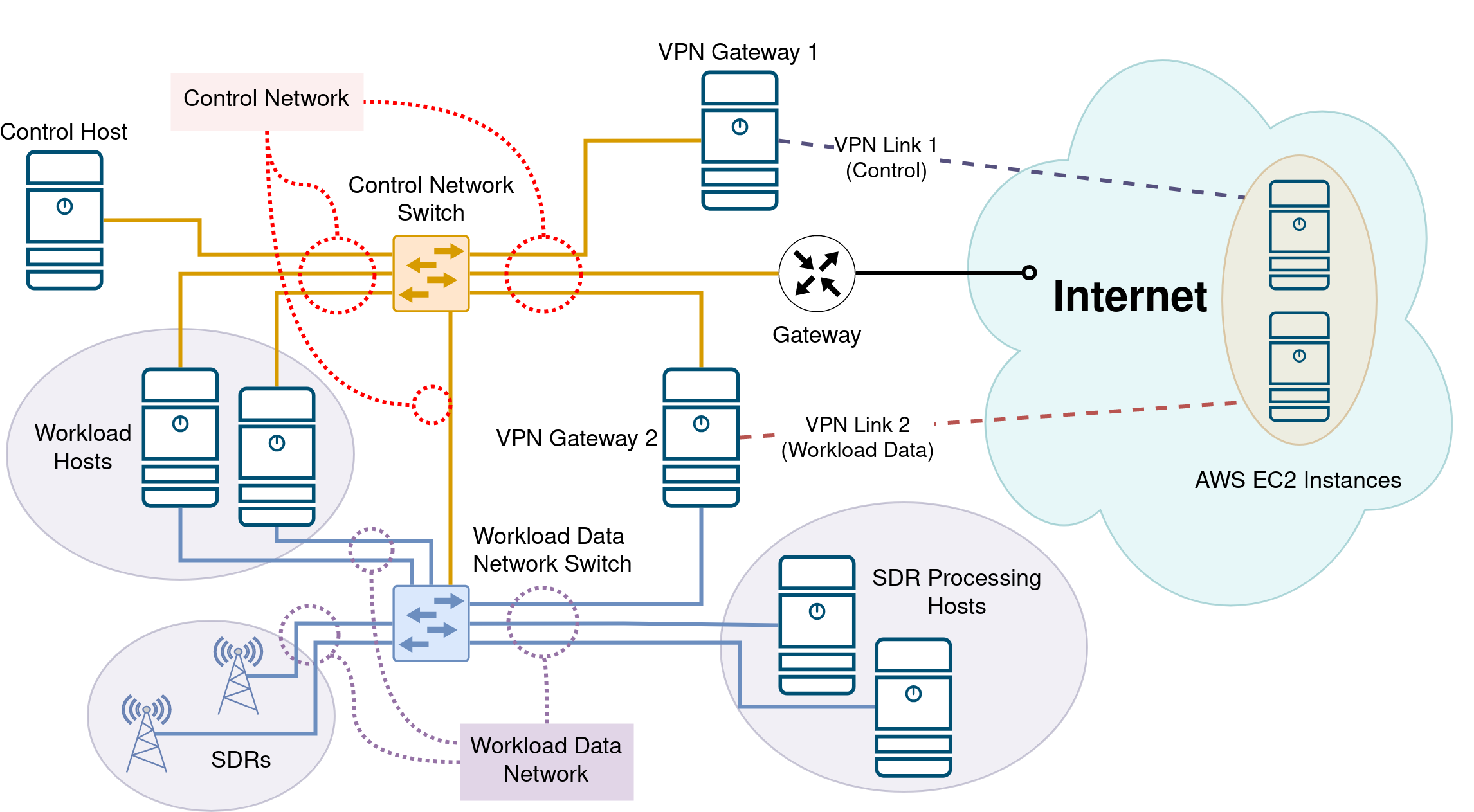}
    \caption{Network structure assumed by Ainur.}\label{fig:network}
\end{figure*}

Ainur is designed as an end-to-end wireless network testbed management framework, fully flexible in terms of the communication network stack, computation hosts, and the distributed application deployed on top.
The core goal of the framework is to facilitate the creation of the desired communication and computation elements to discover their implications to the performance of end-to-end distributed applications.
Ainur achieves this by offering a Python \ac{API} which is used to describe an end-to-end experiment in a procedural manner, and we plan to eventually provide toolkits for declarative configuration of experiments.

Conceptually, in Ainur, an experiment is decomposed into a layered structure, consisting of
\begin{enumerate*}[itemjoin={{; }}, itemjoin*={{; and }}]
    \item the distributed application (workload)
    \item computation hosts
    \item communication networks connecting the hosts
\end{enumerate*}.
In \cref{fig:overview}, the workload is represented by a client-server process pair.
These could correspond, for instance, to the emulation of an inverted pendulum and a matching controller.
However, Ainur makes no assumptions about the nature of the workloads deployed on the framework, and virtually any process or combination of processes can be used.

Hosts correspond to either bare-metal machines or cloud instances on \ac{AWS} \ac{EC2}.
Users are free to combine and interconnect these in any configuration.
Ainur can provision wired networks over Ethernet, and supports a number of software-defined wireless communication stacks, currently including WiFi, 4G \ac{LTE}, and 5G.
To realize these various wireless communication protocols, it is assumed that the testbed is equipped with \ac{USRP} \ac{SDR}s and some computation hosts dedicated to signal processing and/or radio management.

To collect data from workloads, Ainur configures both a shared, distributed, storage location which all processes can reach, as well as a logging service which automatically captures structured and unstructured data from the standard output of workload processes.
The logging service additionally collects data from the other two layers, and the resulting dataset could, for instance, be analyzed to relate the performance of the control loop to the quality of the wireless link. 

\subsection{Ainur Software Stack}

To realize our vision for an automated, flexible, cloud-native, and workload-agnostic framework for cloud and edge computing experimentation, we built Ainur on top of a combination of well-established tools and frameworks.
Below we briefly touch on the most important of these:

\begin{description}
    \item[Python:]
    Ainur is built on-top of Python 3.8.
    We chose this language for its flexibility, ease of prototyping, and for the extensive ecosystem of third-party cloud-native frameworks and libraries.

    \item[Ansible:]
    One of the ``cloud-native frameworks'' mentioned above, Red Hat Ansible is a powerful Python framework for bare-metal configuration, provisioning, and automation.
    We use it extensively in Ainur for configuration of network interfaces and services on managed hosts.

    \item[Containers:]
    We leverage Docker containers to great extent for the virtualization and orchestration of workloads, as well as the encapsulation and management of complex network configurations.
    See \cref{ssec:containers} for more details.    

    \item[\ac{AWS} and \texttt{boto3}:]
    Ainur employs the Amazon \texttt{boto3} library for Python to directly interface with \ac{AWS} \ac{EC2} and deploy, configure, and manage remote cloud instances.
    
    \item[\ac{OAI}:]
    \ac{OAI} is an open-source project that implements 3GPP technology on general purpose \texttt{x86} computing hardware and off-the-shelf \acp{SDR} like the \ac{USRP}~\cite{KALTENBERGER2020107284}. 
    Ainur can deploy, configure, and manage \ac{OAI} software components that implement 4G \ac{LTE} and 5G \ac{NR}.
    
    \item[Mango Communications 802.11:]
    Finally, the Mango Communications project implements real-time 802.11 (WiFi) MAC and PHY in Xilinx \acp{FPGA}.
    It can be used on a variety of hardware platforms including \ac{USRP} \acp{SDR}, and is employed in Ainur to provision end-to-end WiFi links.
\end{description}

\subsection{Main Software Components}

Ainur follows a layered architecture which closely mimics the conceptual layers of the \acs{TCP}/\acs{IP} stack.
Components are deployed in bottom-up order, starting with the establishment of physical links, through the establishment of \ac{IP} connectivity and deployment of links to the cloud, and ending with the distribution and initialization of workloads on top of a container orchestration layer.
The architectural modules of the framework can broadly be classified according to the below categories:

\begin{description}[]
    \item[Configuration Layer:]
    The lowest layer of Ainur, it handles parsing of configuration files describing experimental scenarios.
    Optional, as it is only needed when Ainur is running experiments described in a declarative manner.

    \item[Logging layer:]
    This layer handles logging across all layers of the system to a central repository.
    Concretely, it manages the configuration and lifecycle of a Fluentd server to which any component in the network can send logs.
    Currently, the two main components relying on this scaffolding are the physical layer and the workloads.
    
    \item[Physical Layer:] 
    Creates and deploys the underlying physical connections of the workload data network.
    This layer interacts with hardware such as managed switches and \acp{SDR} (and their associated computation hosts) to create links between the desired workload hosts.
    These links correspond to ethernet, WiFi, and even an entire software-defined 4G \ac{LTE} and 5G networks.

    \item[Cloud Layer:]
    Handles integration with cloud services (currently, \ac{AWS} \ac{EC2}). 
    Only deployed in the case of an experiment requiring cloud instances, it manages the instantiation and configuration of remote cloud resources.

    \item[\ac{IP} Layer:]
    Configures and establishes \ac{IP} layer connectivity of packets between hosts in the experimental setup, both local and cloud.
    This includes assigning valid \ac{IP} addresses to hosts, establishing \ac{VPN} routes to cloud instances through a pair of pre-configured \ac{VPN} gateways, and configuring routing tables to ensure any two hosts in the workload network can communicate with each other.

    \item[Workload Layer:]
    Finally, this layer deploys, scales, and orchestrates containerized workloads, as well as configures a shared, distributed storage for workloads to store data.
    This components leverages Docker Swarm to spin up and manage workload containers on desired hosts.
    It also establishes overlay networks abstracting away the physical topology and allowing containers to interconnect through dynamically assigned hostnames.
\end{description}

\subsection{Containerization}\label{ssec:containers}

Containers are a key cloud-native technology for the virtualization and sandboxing of arbitrary processes~\cite{merkel2014docker}.
They allow for easy packaging of software in predefined, consistent, and conflict-free execution environments, including all necessary dependencies.
Containers are lightweight and portable compared to \ac{VM}-based virtualization, making them an ideal solution for the distribution, deployment, and orchestration of software in distributed computing environments.
They deploy quickly and are very configurable, while abstracting away the complexities and delays that come with in creating, managing, and moving (potentially huge) \ac{VM} disk images.

Ainur leverages containerization extensively across the framework, in order to
\begin{enumerate*}[itemjoin={{, }}, itemjoin*={{, and }}]
    \item deploy and orchestrate workloads
    \item support a wide spectrum of different physical layer configurations out-of-the-box, as well as allow for easy extension to new ones
    \item support automated collection of logs
\end{enumerate*}.
Workloads in the framework are deployed packaged in Docker containers, orchestrated through \texttt{docker-compose} and Docker Swarm.
This allows the framework to remain mostly agnostic to the nature of the workloads, and thus support a wide ranged of different applications and system architectures.
It also allows for easy deployment to different compute nodes in the network, without having to take into consideration details such as required libraries and versions.

As expressed above, Ainur also leverages containers for the communication stack.
With the advent of \ac{O-RAN} and \acp{SDR}, all components of the wireless network stack can run on general-purpose processors.
They can thus be containerized and distributed across multiple hosts.

As a final point, the automation of logging is another advantage of using containers and an orchestration framework.
Ainur employs the Fluentd unified layer for log collection.
It natively integrates with Docker containers and allows for the automatic collection of text data from the standard output and error streams of processes executing inside containers.
It automatically decouples data sources from containers running on different systems and allows Ainur to collect and classify the logs from any components of the network, whether on the wireless stack or workload.

\subsection{General assumptions}

Ainur makes a number of assumptions about its execution environment.
Apart from generic ones regarding the presence of necessary authentication information for remote access to local and remote hosts and services, the framework assumes a split network architecture such as the one depicted in \cref{fig:network}.
Ainur runs in a dedicated control host and the management plane resides in a physically distinct network from the workload data.
This is a key requirement to be able to reconfigure the physical links in the workload data network without disrupting management traffic.
Also assumed is the existence of \ac{VPN} gateways to the cloud, time synchronization, and configured hostname lookups (possibly using a \ac{DNS} service).

\subsection{Obtaining Ainur}

The framework is released as \acl{FOSS} under a permissive Apache license.
It can be downloaded from the Ainur repository~\cite{ainur:github} of the {KTH-EXPECA} organization on GitHub.
Given the software's complexity, the repository also includes detailed documentation on its requirements, configuration, deployment, and execution, as well as links to external resource containing information and guides about related concepts and services.

\begin{figure*}
    \centering
    \includegraphics[width=.8\textwidth]{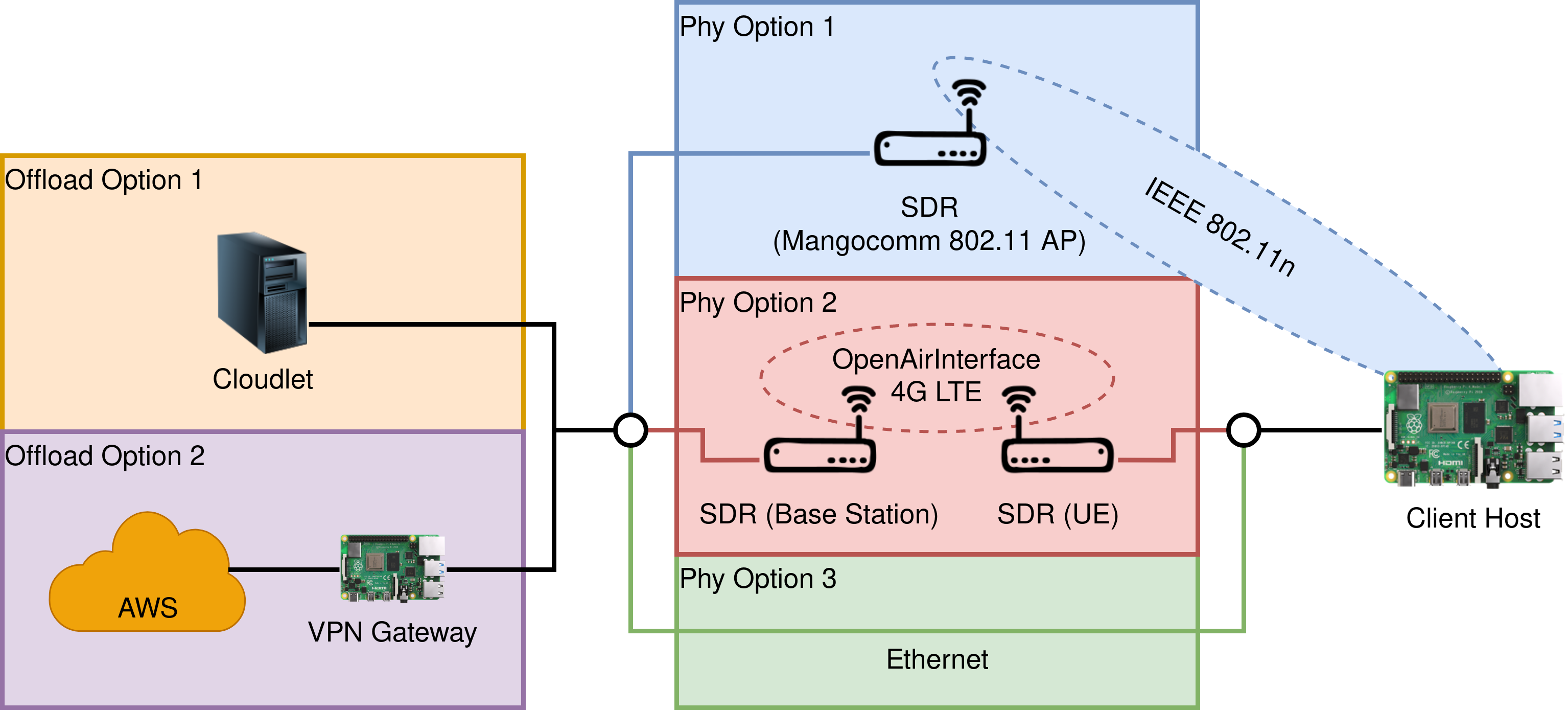}
    \caption[caption]{%
        The possible workload offloading setups and physical layer configurations used in the demo.
        Not pictured, in Phy Option 2:
        \begin{enumerate*}[itemjoin={{; }}, itemjoin*={{; and }}]
            \item \ac{SDR} processing hosts 
            \item additional base-station host for the Core Network and \ac{eNodeB}
        \end{enumerate*}.
    }\label{fig:democonfigs}
\end{figure*}

\begin{figure}
    \centering
    \begin{subfigure}{\columnwidth}
        \centering
        \includegraphics[height=8em]{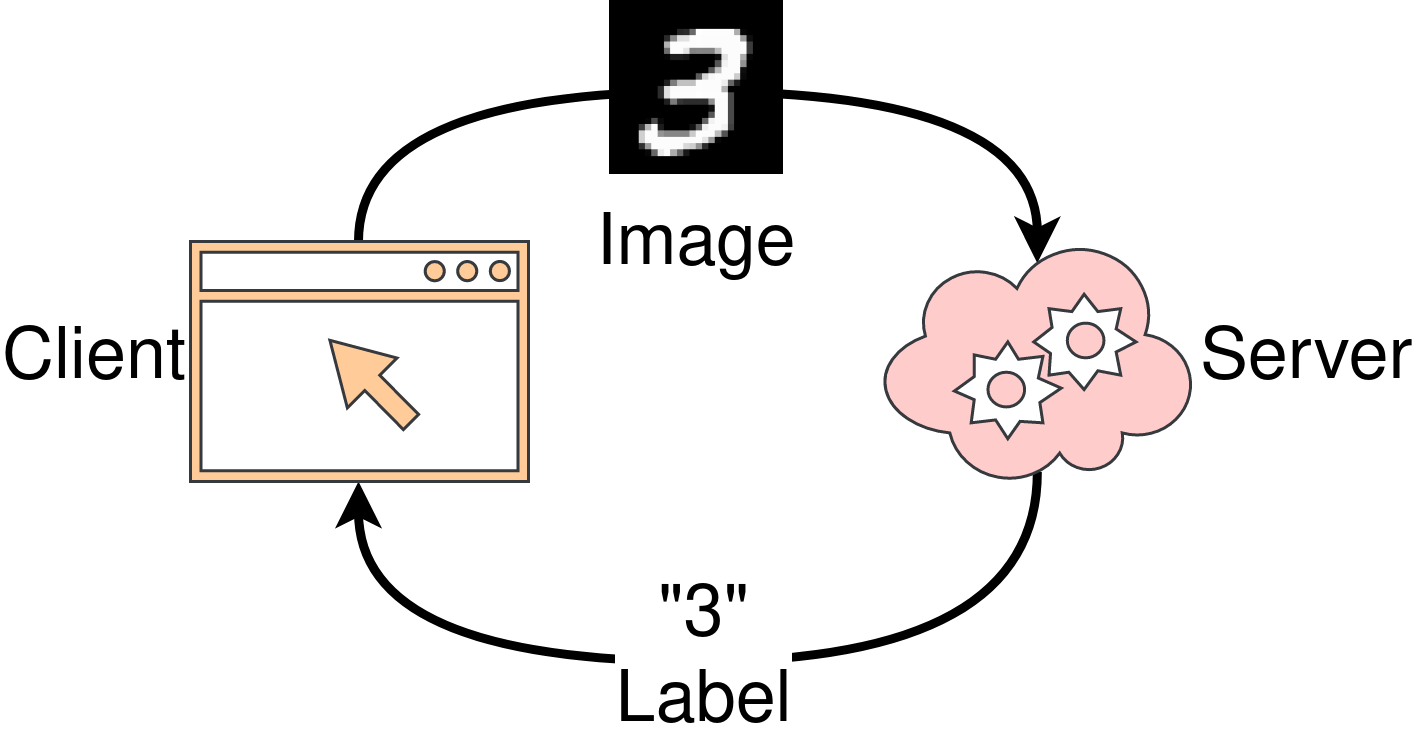}
        \caption{Workload 1: \acs*{MNIST} image classifier}\label{fig:wkld:mnist}
    \end{subfigure}%
    \vspace{1em}
    \begin{subfigure}{\columnwidth}
        \centering
        \includegraphics[height=8em]{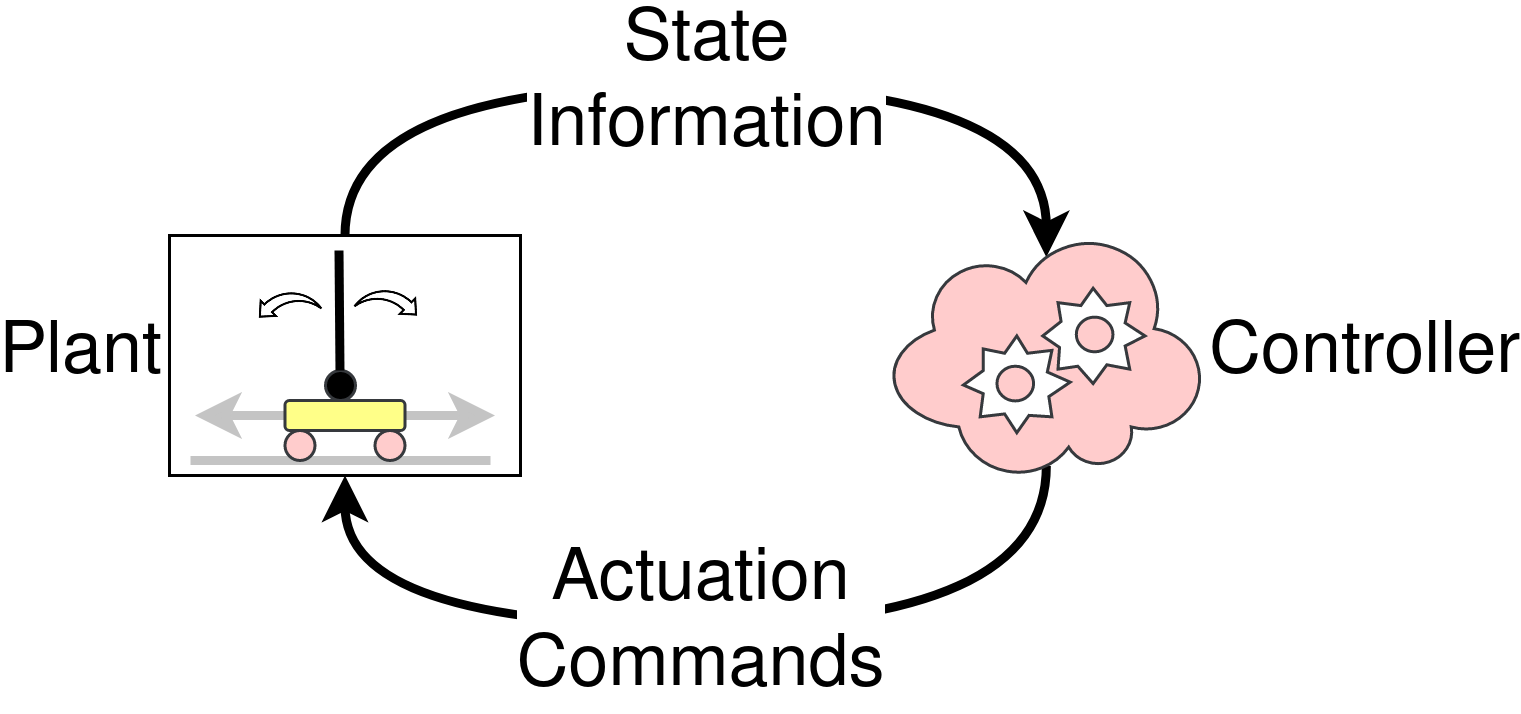}
        \caption{Workload 2: inverted pendulum \acs*{NCS}}\label{fig:wkld:ncs}
    \end{subfigure}
    \caption{Demonstration workloads}\label{fig:wkld}
\end{figure}

\section{Demo}\label{sec:demo}

In this demo, we will show the flexibility of Ainur for running end-to-end experimental workloads on both the Edge and the Cloud.
This will be done interactively, and members of the audience will be invited to propose testbed configurations.

\Cref{fig:democonfigs} illustrates the possible configurations for the testbed.
Workloads are deployed on client hosts, and computation is offloaded either to an edge server (cloudlet), or to \ac{AWS} \ac{EC2} instances on the cloud.
Communication between the client- and server-sides of the workload will occur over one of three possible physical link layer setups:
\begin{description}
    \item[WiFi:] \iac{SDR} is configured as an IEEE 802.11n access point, and client hosts connect to it using on-board WiFi.
    \item[4G \ac{LTE}:] \iac{SDR} is configured as an 4G \ac{LTE} base station, and another is configured as an 4G \ac{LTE} \ac{UE}.
    Together these radios act as an \ac{LTE} bridge between the client-side and the server-side of the network, and all client-server traffic is routed through them.
    \item[Ethernet:] connects everything through plain ethernet.
\end{description}

The number of workload instances (and therefore client hosts) deployed in each execution of this demonstration will range from \numrange[]{1}{10}.
Workloads deployed to the cloud will be able to target any of \ac{AWS}'s datacenters.

We will employ two different workloads for this demonstration, illustrated in \cref{fig:wkld}.
The first of these (\cref{fig:wkld:mnist}) consists of a proof-of-concept web application which implements a simple classifier to identify hand-drawn digits from the widely-used \ac{MNIST} dataset~\cite{mnist}.
The application consists of a client with a web interface, and \iac{HTTP} server which responds to requests from the client and performs the actual image recognition.
This workload is intended to showcase in an interactive manner the effects of placing computation at different points of the network, and how Ainur simplifies these deployments.

The second workload (\cref{fig:wkld:ncs}) corresponds to \iac{NCS} balancing an inverted pendulum, implemented on a software framework for the emulation of \acp{NCS} using cloud-native technologies~\cite{cleave}.
It consists of an emulation of the physical inverted pendulum system plant and a software-implemented proportional-differential controller.
These components communicate with each other over the \ac{UDP}.
This workload will be used to showcase the utility of Ainur for automating the execution of batches of experiments potentially including multiple different clients, servers, and physical layers.

\subsection{Testbed Setup}

This demonstration will be performed on a testbed consisting of
\begin{enumerate*}[itemjoin={{; }}, itemjoin*={{; and finally }}]
    \item \num{10} Raspberry Pi 4 Model B boards, acting as client-side workload hosts
    \item a Raspberry Pi 4 Model B acting as \ac{VPN} gateway for the workload network
    \item a Raspberry Pi 4 Model B acting as \ac{VPN} gateway for the management network, as well as hosting the necessary \ac{NTP} and \ac{DNS} server software
    \item a \texttt{i386} workstation, acting as an edge-side workload host
    \item a configurable number of  \ac{AWS} \ac{EC2} cloud instances acting as cloud servers
    \item a separate \texttt{i386} workstation hosting the Fluent server and on which Ainur is deployed as well.
\end{enumerate*}
These nodes are interconnected using a combination of managed switches, \acp{SDR}, and \ac{VPN} gateways; please refer to \cref{fig:network} for an architectural overview of this setup.

\subsection{Demo Procedure}

\subsubsection{Workload 1}

\begin{figure}
    \centering
    \includegraphics[width=.65\columnwidth]{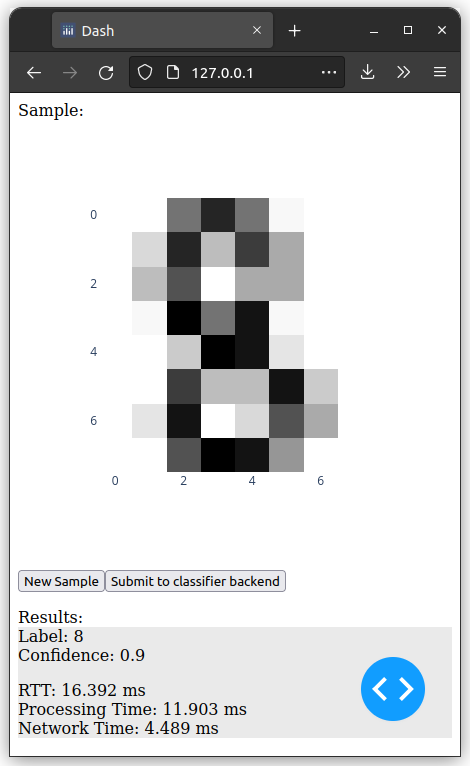}
    \caption{Screenshot of the user interface of demo workload 1. After deciding on placement of the compute backend and the type of physical layer connecting client and backend, participants will use this interface to interact with the system.}\label{fig:mockup}
\end{figure}

Participants will be asked to choose
\begin{enumerate*}[itemjoin={{; }}, itemjoin*={{; and }}]
    \item a location to offload computation to (local (i.e.\ no offloading), edge, or any \ac{AWS} datacenter)
    \item a physical layer for the first hop of the network (WiFi, 4G \ac{LTE}, or Ethernet). 
\end{enumerate*}
Through the Ainur command line, we will deploy a single client-server pair according to the specifications.
Next, participants will be able to access the client web interface and interact with the server by requesting classification of images from the \ac{MNIST} database.
The server will respond to these, and the assigned labels will be visible on the web interface together with timing statistics for each request, see \cref{fig:mockup}.

\subsubsection{Workload 2}

Participants will be asked to specify the full, arbitrarily complex, experimental scenario, including the number of clients, physical layers (single or multiple, and which clients are on each physical link), and offloading configuration (number of instances on the edge vs.\ on the cloud, which \ac{AWS} datacenters to deploy to).
Configuration will be specified through a \acs{YAML} file, which will then be parsed by Ainur for the automatic execution of the scenario.
\section{Conclusion}\label{sec:conclusion}

In this paper, we have introduced the Ainur framework for repeatable end-to-end testbed automation in the context of wireless networking and edge computing research.
The framework simplifies the execution and verification of end-to-end experimentation in these testbeds by automating the
\begin{enumerate*}[itemjoin={{; }}, itemjoin*={{; and }}]
    \item establishment of physical links between hosts, including the configuration of complex wireless systems such as 4G \ac{LTE} and 5G
    \item provisioning of and connection to remote cloud instances
    \item initialization of \ac{IP} layer connectivity between hosts
    \item collection of logs and data
    \item deployment, scaling, and lifecycle management of containerized processes
\end{enumerate*}.
We have described its general architecture, which follows a layered design mimicking the network stack layers the framework directly interacts with, as well as the underlying assumptions about its deployment environment and specific requirements for its deployment.

Finally, we have outlined a demonstration which showcases the flexibility and power of the framework by deploying two different workloads to our testbed.
We show how Ainur deployment of these to different points in the network, as well as how the framework makes possible repeatable, large-scale experimentation in distributed systems.

We believe our framework represents an important step towards repeatable, replicable, yet low-access barrier end-to-end wireless testbed experimentation.
It has been released as \ac{FOSS} and can be found on GitHub~\cite{ainur:github}.

\printbibliography{}

@inbook{
Cosmos1,
author = {Raychaudhuri, Dipankar and Seskar, Ivan and Zussman, Gil and Korakis, Thanasis and Kilper, Dan and Chen, Tingjun and Kolodziejski, Jakub and Sherman, Michael and Kostic, Zoran and Gu, Xiaoxiong and Krishnaswamy, Harish and Maheshwari, Sumit and Skrimponis, Panagiotis and Gutterman, Craig},
title = {Challenge: COSMOS: A City-Scale Programmable Testbed for Experimentation with Advanced Wireless},
year = {2020},
isbn = {9781450370851},
publisher = {Association for Computing Machinery},
address = {New York, NY, USA},
doi = {10.1145/3372224.3380891},
booktitle = {Proceedings of the 26th Annual International Conference on Mobile Computing and Networking},
articleno = {14},
numpages = {13}
}

@INPROCEEDINGS{cosmos2,
  author={Yu, Jiakai and Chen, Tingjun and Gutterman, Craig and Zhu, Shengxiang and Zussman, Gil and Seskar, Ivan and Kilper, Daniel},
  booktitle={2019 Optical Fiber Communications Conference and Exhibition (OFC)}, 
  title={COSMOS: Optical Architecture and Prototyping}, 
  year={2019},
  volume={},
  number={},
  pages={1-3},
  doi={}}

@inproceedings{powder,
  author =       "Joe Breen and Andrew Buffmire and Jonathon Duerig and Kevin
                  Dutt and Eric Eide and Mike Hibler and David Johnson and
                  Sneha Kumar Kasera and Earl Lewis and Dustin Maas and Alex
                  Orange and Neal Patwari and Daniel Reading and Robert Ricci
                  and David Schurig and Leigh B. Stoller and Van der Merwe,
                  Jacobus and Kirk Webb and Gary Wong",
  title =        "{POWDER}: Platform for Open Wireless Data-driven Experimental
                  Research",
  booktitle =    "Proceedings of the 14th International Workshop on Wireless
                  Network Testbeds, Experimental Evaluation and
                  Characterization (WiNTECH)",
  year =         2020,
  month =        sep,
  doi =          "10.1145/3411276.3412204",
}

@INPROCEEDINGS{cloudraft,
  author={Moorthy, Sabarish Krishna and Lu, Chencheng and Guan, Zhangyu and Mastronarde, Nicholas and Sklivanitis, George and Pados, Dimitris and Bentley, Elizabeth Serena and Medley, Michael},
  booktitle={2022 IEEE 19th Annual Consumer Communications   Networking Conference (CCNC)},
  title={CloudRAFT: A Cloud-based Framework for Remote Experimentation for Mobile Networks},
  year={2022},
  volume={},
  number={},
  pages={1-6},
  doi={10.1109/CCNC49033.2022.9700510}
}

@article{omf,
author = {Rakotoarivelo, Thierry and Ott, Maximilian and Jourjon, Guillaume and Seskar, Ivan},
title = {OMF: A Control and Management Framework for Networking Testbeds},
year = {2010},
issue_date = {January 2010},
publisher = {Association for Computing Machinery},
address = {New York, NY, USA},
volume = {43},
number = {4},
issn = {0163-5980},
doi = {10.1145/1713254.1713267},
journal = {SIGOPS Oper. Syst. Rev.},
month = {jan},
pages = {54-59},
numpages = {6}
}

@INPROCEEDINGS{orbit,
  author={Ott, M. and Seskar, I. and Siraccusa, R. and Singh, M.},
  booktitle={First International Conference on Testbeds and Research Infrastructures for the DEvelopment of NeTworks and COMmunities},
  title={ORBIT testbed software architecture: supporting experiments as a service},
  year={2005},
  volume={},
  number={},
  pages={136-145},
  doi={10.1109/TRIDNT.2005.27}
}

@ONLINE{maas, 
  author={Canonical},
  title={Metal-As-A-Service (MAAS)},
  url={https://maas.io}
}

@INPROCEEDINGS{DrexelGrid,
author={Dandekar, Kapil R. and Begashaw, Simon and Jacovic, Marko and Lackpour, Alex and Rasheed, Ilhaan and Rey, Xaime Rivas and Sahin, Cem and Shaher, Sharif and Mainland, Geoffrey},
booktitle={2019 16th Annual IEEE International Conference on Sensing, Communication, and Networking (SECON)},
title={Grid Software Defined Radio Network Testbed for Hybrid Measurement and Emulation},
year={2019},
volume={},
number={},
pages={1-9},
doi={10.1109/SAHCN.2019.8824901}
}

@article{apt,
    author      = "Robert Ricci and Gary Wong and Leigh Stoller and Kirk Webb and Jonathon Duerig and Keith Downie and Mike Hibler",
    title       = "Apt: A Platform for Repeatable Research in Computer Science",
    journal     = "{ACM} {SIGOPS} Operating Systems Review",
    year        = 2015,
    month       = jan,
    volume      = 49,
    number       = 1
}

@article{mnist,
  author  = {Deng, Li},
  journal = {IEEE Signal Processing Magazine},
  title   = {The MNIST Database of Handwritten Digit Images for Machine Learning Research [Best of the Web]},
  year    = {2012},
  volume  = {29},
  number  = {6},
  pages   = {141-142},
  doi     = {10.1109/MSP.2012.2211477}
}

@inproceedings{cleave,
  author    = {Olgu\'{\i}n Mu\~{n}oz, Manuel and Roy, Neelabhro and Gross, James},
  title     = {{CLEAVE}: Scalable and Edge-Native Benchmarking of Networked Control Systems},
  year      = {2022},
  isbn      = {9781450392532},
  publisher = {Association for Computing Machinery},
  address   = {New York, NY, USA},
  doi       = {10.1145/3517206.3526272},
  booktitle = {Proceedings of the 5th International Workshop on Edge Systems, Analytics and Networking},
  pages     = {37-42},
  numpages  = {6},
  location  = {Rennes, France},
  series    = {EdgeSys '22}
}

@article{merkel2014docker,
  title   = {Docker: lightweight linux containers for consistent development and deployment},
  author  = {Merkel, Dirk and others},
  journal = {Linux journal},
  volume  = {2014},
  number  = {239},
  pages   = {2},
  year    = {2014}
}

@misc{ainur:github,
  title     = {Ainur GitHub Repository},
  url       = {https://github.com/KTH-EXPECA/Ainur},
  journal   = {GitHub},
  publisher = {EXPECA group at KTH}
}

@article{KALTENBERGER2020107284,
title = {OpenAirInterface: Democratizing innovation in the 5G Era},
journal = {Computer Networks},
volume = {176},
pages = {107284},
year = {2020},
issn = {1389-1286},
doi = {https://doi.org/10.1016/j.comnet.2020.107284},
url = {https://www.sciencedirect.com/science/article/pii/S1389128619314410},
author = {Florian Kaltenberger and Aloizio P. Silva and Abhimanyu Gosain and Luhan Wang and Tien-Thinh Nguyen}
}
\end{document}